\documentclass[aps,prb,reprint,floatfix,longbibliography]{revtex4-2}
\usepackage{amsmath,amssymb}
\usepackage{graphicx,booktabs}

\newcommand{\chiz}{\chi_0}
\newcommand{\Leff}{\Lambda_{\rm eff}}

\newcommand{\Reff}{R_{\rm eff}}
\newcommand{\um}{\,$\mu$m}

\begin{document}

\title{Magnetic Susceptibility of Superconducting Films in the Meissner-London Regime}
\author{Ruslan Prozorov}
\email{prozorov@ameslab.gov}
\affiliation{Ames National Laboratory, Ames, Iowa 50011, USA}
\affiliation{Department of Physics and Astronomy, Iowa State University, Ames, Iowa 50011, USA}
\date{14 October 2026}

\begin{abstract}
Extracting the London penetration depth, $\lambda(T)$, from the measured magnetic susceptibility, $\chi(T)$, of a superconducting platelet in a perpendicular magnetic field requires knowing the function $\chi(\lambda)$ computed for the actual sample geometry. Thin films are increasingly important in quantum computing and superconducting electronics, most of which operate at ultra-low excitation fields in the vortex-free regime. Understanding their electromagnetic response in the Meissner--London (finite-$\lambda$, no vortices) state is therefore essential for extracting intrinsic properties such as the London penetration depth and superfluid phase stiffness. While numerous prior works have addressed different aspects of this problem, here we focus on providing an analytical semi-empirical representation of $\chi(\lambda)$ based on the 2D sheet-current model and establishing the limits of its applicability using two independent full 3D numerical solutions: finite-element modeling and a volume-current integral equation. As a result, we provide $\chi(\lambda)$ for a disc of radius $a$ and thickness $d$ in closed form, valid for all $\lambda$ and for aspect ratios $\delta\equiv d/(2a)\le0.05$, and accurate to $0.1\%$. The two-dimensional sheet description with Pearl length $\Lambda=2\lambda^2/d$ is adequate for $\delta \lesssim10^{-4}$ and $\lambda \gtrsim 4d$. Thicker films require the full closed form, which includes the three-dimensional corrections. Using the same numerical analysis, we provide an empirical formula that maps rectangular samples---the geometry most commonly used in experiments---onto a disc with an ``effective radius'' for subsequent calculation of $\chi(\lambda)$, and formulate a complete calibration protocol for whole-sample susceptibility measurements.
\end{abstract}

\maketitle

\section{Introduction}
Small-amplitude (vortex-free) Meissner--London susceptibility in a perpendicular magnetic field is an established probe for extracting the London penetration depth, $\lambda$, and thus the superfluid density \cite{topping2019}. We consider a thin disc of radius $a$ and thickness $d$, placed in a uniform applied magnetic field, $\mathbf H=H\hat z$. In the London gauge, the vector potential $\mathbf A$ satisfies $\nabla\times\nabla\times\mathbf A+\mathbf A/\lambda^2=0$ inside the sample and $\nabla\times\nabla\times\mathbf A=0$ outside. We focus on linear magnetic susceptibility, $\chi=m/(VH)$, where $m$ is the total magnetic moment and $V=\pi a^2d$ is the sample volume, and we characterize the geometry by the aspect ratio, defined as the full thickness over the diameter, $\delta\equiv d/(2a)$. Radial positions are given in units of the radius, $x=r/a$.

Two classes of experiments use perpendicular-field susceptibility, and each poses a distinct electromagnetic problem.

In a conventional \emph{whole-sample} measurement, the entire sample is placed in a nearly uniform magnetic field, and the measured signal is proportional to its total magnetic moment, $m$ \cite{prozorov2006,topping2019,moirangthem2026,moirangthem2026b}. In thin films, this response is governed by the demagnetizing factor, dominated by the sample edges, and controlled by the competition between the Pearl screening length $\Lambda=2\lambda^2/d$~\cite{pearl1964} and the lateral sample size $a$.

The other class of $\chi$ measurements frequently used for thin films comprises two-coil mutual-inductance~\cite{fiory1988,turneaure1996,claassen1997,turneaure1998,lemberger2007,zhang2021,harvey2025} and miniature scanning SQUID susceptometry~\cite{kogan2003,huber2008,bert2011,kirtley2012,bert2012,low2021}, where a film much larger than the pickup/drive coils is excited locally. Unless the film is comparable in size to the coils~\cite{zhang2021}, the lateral sample dimensions drop out, the relevant screening length competes with the coil radius and probe height, and the analysis laterally assumes an infinite film.

This paper addresses the whole-sample problem, in which a measurement returns a quantity proportional to $\chi(T)$ with an unknown apparatus calibration constant $G$. Extracting the intrinsic London penetration depth, $\lambda(T)$, then requires knowing the function $\chi(\lambda,d,a)$ for the actual sample geometry.

This problem has been examined in many previous works from different perspectives and under various approximations. Most treatments rely on the 2D sheet-current approximation, in which the film is replaced by an infinitely thin sheet carrying the thickness-integrated current density, with Pearl's screening length derived for a current uniform across the thickness $d$. Here, we use two independent full 3D solutions to test and establish the validity bounds of these assumptions.

When $\lambda\to0$ and $\delta\to0$, the diamagnetic susceptibility of a thin disc is analytic, with magnitude $\chiz\equiv|\chi(\lambda\to0,\delta\to0)|=8a/(3\pi d)=4/(3\pi\delta)$~\cite{mikheenko1993,clem1994}. Brandt's treatment of the general linear response of thin discs~\cite{brandt1994b,brandt1994c} provides the strong-screening asymptote and a tabulated pole expansion, although it was formulated for ac losses and, to our knowledge, has not been systematically used to invert vortex-free London penetration depth data. Square and rectangular sheets were treated by Brandt in subsequent works \cite{brandt1995,brandt2005}. Chen \emph{et al.} computed the ideal ($\lambda=0$) magnetic susceptibility of rectangular films of arbitrary aspect ratio~\cite{chen2008,chen2010}, solved the thin strip at finite $\Lambda$ \cite{chen2008b}, and inverted the square-film response to obtain $\lambda(T)$ \cite{chen2014}. Their low-$\Lambda$ empirical law for the strip, $1-|\chi|/\chiz\approx5.2\,\Lambda/w$ ($w$ being the half-width), omits the logarithmic factor produced by the edge singularity: first-order perturbation in the kinetic energy gives the same functional form as for the disc, $(2/\pi)(\Lambda/w)\ln(Cw/\Lambda)$, and a strip version of our sheet solver gives $C\simeq5.3$. The same law, with $C=16.2/\pi=5.16$, is contained in Brandt's high-frequency asymptote for the Ohmic strip \cite{brandt1994a} continued to the London regime, exactly as for the disc. The purely linear law is therefore adequate near $\Lambda/w=10^{-3}$, where it was fitted (within $4\%$), but underestimates the susceptibility deficit by $25\%$ at $\Lambda/w=10^{-4}$ and by $38\%$ at $10^{-5}$. For thicker samples, the ideal ($\lambda=0$) limit is known for arbitrary aspect ratios \cite{brandt1996,brandt1998,brandt2001}. For thick London cylinders, finite-$\lambda$ empirical formulas based on full 3D numerical solutions are given in our previous work~\cite{prozorov2021}.

In this work, we provide a closed-form expression for $|\chi(\lambda,d,a)|$ of a circular disc valid for all $\lambda$ and across the full range of practical film thicknesses used in modern experiments ($\delta\le0.05$). We use two independent full 3D Meissner--London solvers to determine its accuracy and the boundaries of the 2D sheet regime.

\section{Three-dimensional solutions}\label{sec:methods}

The full 3D axisymmetric Meissner--London problem was solved by two independent numerical methods:

(1) Finite-element analysis using COMSOL Multiphysics in an axisymmetric formulation for a disc of radius $a=3$~mm and thicknesses ranging from $d=4$~nm to $300$\um\ ($\delta=6.7\times10^{-7}$ to $0.05$). The adaptive mesh was refined near the rim and sample surfaces to fully resolve the spatial distributions of magnetic fields and supercurrents inside and around the disc.

\begin{figure}[tbh]
\includegraphics[width=8.6cm]{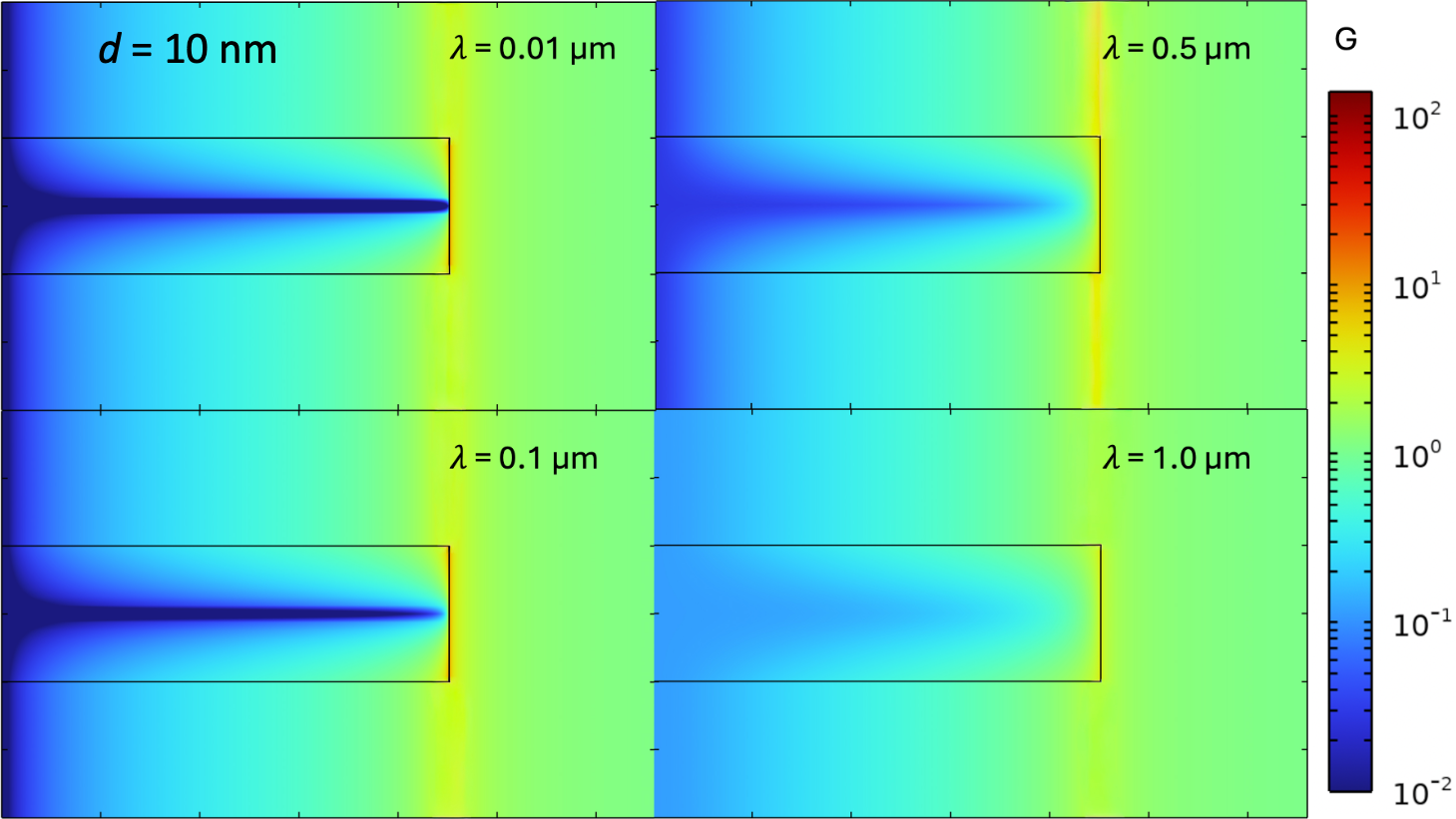}
\caption{Cross-sectional distribution of the $z$-component of the magnetic induction, $B_z$ (in G, for an applied field $\mu_0 H=1$~G), near the rim of a $d=10\,\textrm{nm}$, $a=3\,\textrm{mm}$ superconducting disc for four values of the London penetration depth $\lambda$ indicated in the panels. To visually resolve the distribution across the thickness, the vertical $z$-axis is stretched by a factor of $10^5$ relative to the radial $r$-axis.}
\label{fig:COMSOL_Bz}
\end{figure}

Figure~\ref{fig:COMSOL_Bz} shows the cross-sectional distribution of axial magnetic induction $B_z(r,z)$ in a thin film ($d=10\,\textrm{nm}$, $a=3\,\textrm{mm}$) for four values of the London penetration depth $\lambda$ indicated in the panels. To make the variation across the $10\,\textrm{nm}$ thickness visible alongside the radial penetration from the film's edge, the vertical $z$-scale is expanded by a factor of $10^5$ relative to the radial $r$-scale. The minimum of $B_z$ at the midplane reflects the $\cosh(z/\lambda)$ profile of the vector potential and of the current across the film (Sec.~\ref{sec:model}).

\begin{figure}[tbh]
\includegraphics[width=8.4cm]{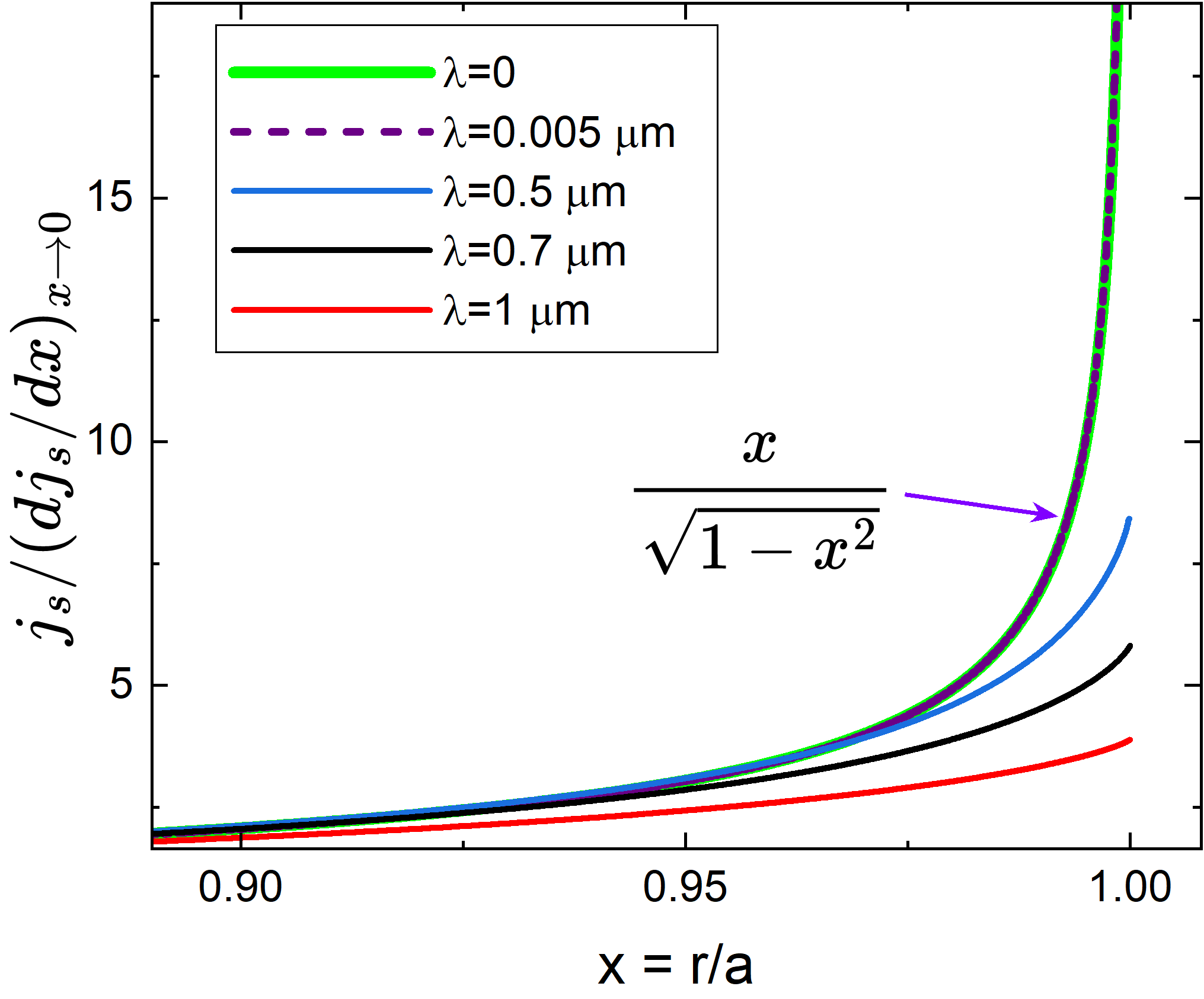}
\caption{Thickness-integrated sheet current density, $j_s(x)$, normalized by the central slope of the ideal profile, $(dj_s/dx)_{x\to0}=4H/\pi$, as a function of $x=r/a$ near the rim of a $d=10\,\textrm{nm}$, $a=3\,\textrm{mm}$ disc; the curves end at $x=0.9999$. For $\lambda \to 0$ ($\lambda=0.005$\um, dashed curve), the numerical solution coincides with the exact 2D ideal-screening profile $x/\sqrt{1-x^2}$ \cite{mikheenko1993,clem1994,brandt1994b} (solid green curve).}
\label{fig:COMSOL_j}
\end{figure}

To verify the numerical accuracy, we integrated the azimuthal current density over the film thickness to obtain the sheet current density, $j_s(x)=\int_{-d/2}^{d/2} j(x,z)\,dz$. In the 2D ideal-screening limit ($\delta\to0$, $\lambda=0$), the magnitude of the sheet current is analytic, $j_s(x)=(4H/\pi)\,x/\sqrt{1-x^2}$ \cite{mikheenko1993,clem1994,brandt1994b}, so that its initial slope at the center is $(dj_s/dx)_{x\to0}=4H/\pi$. Figure~\ref{fig:COMSOL_j} compares a set of numerically computed $j_s(x,\lambda)$ curves against this analytic profile (green curve). Our smallest penetration depth calculation, $\lambda=5\,\textrm{nm}$ (purple dashed curve), lies directly on top of the $\lambda=0$ analytic curve. With increasing $\lambda$, screening weakens and $j_s(x,\lambda)$ rounds off near the rim $x\to1$ as expected. While $\lambda$ cannot be set identically to zero inside the finite-$\lambda$ London PDE solver in COMSOL, an independent perfect-shielding boundary-value problem ($\mathbf B\cdot\hat n=0$ on the sample surface, which is numerically much simpler) was solved on the identical mesh for every geometry to verify that the $\lambda \to 0$ extrapolation of the London solution always converges to the exact $\lambda=0$ limit.

(2) The second, independent 3D method solves the volume-current integral equation for $j(r,z)$ directly~\cite{brandt1998}: the $r$--$z$ cross-section of the disc is partitioned into $N_r\times N_z$ coaxial rectangular rings of uniform current density, coupled via ring--ring mutual inductances computed from the coaxial-filament formula integrated over the ring cross-sections~\cite{grover1946,brandt1998}, with the mesh graded geometrically toward the rim $r=a$ and both surfaces $z=\pm d/2$. No exterior air mesh is required. Across 30 benchmark $(\lambda,d)$ parameter points spanning the full regime of interest, the two 3D methods agree to better than $0.07\%$.

Having established two mutually consistent 3D solutions, we use them as the exact reference to quantify when the 2D sheet-current approximation breaks down and to constrain our closed-form model across all $\lambda$ and $\delta\le0.05$.

\textit{Effect of sharp edges.} Mikitik~\cite{mikitik2025} recently suggested that sharp corners of a sample with rectangular cross-section add a singular contribution $\propto\lambda^{2/3}$ to the low-$\lambda$ susceptibility deficit, which would dominate the standard linear-in-$\lambda$ term in thin samples. Our rim-resolved 3D solutions do not show such a term. At a right-angle, the ideal ($\lambda=0$) surface magnetic field diverges as $\rho^{-1/3}$ with distance $\rho$ from the corner; consequently, the surface integral $\oint H_t^2\,dS$ that determines the first-order ($\propto\lambda$) London correction converges, and the corner region $\rho\lesssim\lambda$ where the Meissner singular field is cut off contributes only at order $\int_0^\lambda \rho^{-2/3}\rho\,d\rho \propto \lambda^{4/3}$. Direct fits of $|\chi|=|\chiz|-A\lambda+B\lambda^p$ to our 3D results for a disc with $\delta=0.05$ return $p\simeq1.5$. Forcing $p=2/3$ degrades the fit residuals fifteen-fold, and evaluating the $\lambda^{2/3}$ term with the coefficients quoted in Ref.~\cite{mikitik2025} would exceed the total numerically observed deficit tenfold. We attribute the spurious $\lambda^{2/3}$ term in Ref.~\cite{mikitik2025} to the sharp spatial cut-off imposed there. In addition, we explored filleted corners and found no difference, provided the fillet radius is much smaller than the film thickness.

\section{The two-dimensional sheet current model}\label{sec:sheet}

When the film is much thinner than the lateral length scales on which the current varies, only the thickness-integrated sheet current, $j_s(r)=\int_{-d/2}^{d/2} j(r,z)\,dz$, enters the exterior field problem. Integrating London's equation $\mu_0\lambda^2\mathbf j=-\mathbf A$ over the thickness under the assumption that $\mathbf A$ is uniform across $d$ ($d\ll\lambda$) yields the local constitutive law of a Pearl sheet~\cite{pearl1964},
\begin{equation}
j_s=-\frac{2}{\mu_0\Lambda}\,A,\qquad \Lambda=\frac{2\lambda^2}{d},\label{eq:pearl}
\end{equation}
where $A=A_a+A_{j_s}$ is the total azimuthal vector potential $A_\phi(r,z=0)$ in the film plane, comprising the applied-field potential $A_a=\mu_0Hr/2$ and the self-field potential $A_{j_s}$ generated by the sheet current itself. The self-field term follows from the mutual inductance of two coaxial circular filaments \cite{jackson1999},
\begin{equation}
\begin{split}
&A_{j_s}(r)=\frac{1}{2\pi r}\int_0^a\mathcal M(r,r')\,j_s(r')\,dr',\\
&\mathcal M(r,r')=\mu_0\sqrt{rr'}\Big[\Big(\frac2k-k\Big)K(k)-\frac2kE(k)\Big],
\end{split}\label{eq:kernel}
\end{equation}
with $k^2=4rr'/(r+r')^2$ and $K(k)$, $E(k)$ the complete elliptic integrals of the first and second kind. Combining Eqs.~(\ref{eq:pearl}) and (\ref{eq:kernel}) yields a linear Fredholm integral equation of the second kind for $j_s(r)$,
\begin{equation}
\frac{\mu_0\Lambda}{2}\,j_s(r)+\frac{1}{2\pi r}\int_0^a\mathcal M(r,r')j_s(r')\,dr'=-\frac{\mu_0Hr}{2}.\label{eq:sheet}
\end{equation}
Equation~(\ref{eq:sheet}) is the Euler--Lagrange condition minimizing the London free energy functional of the film in the applied field,
\begin{equation}
\begin{split}
\mathcal G[j_s]=\;&\frac12\int_0^a\!\!\int_0^a j_s(r)\,\mathcal M(r,r')\,j_s(r')\,dr\,dr'\\
&+\frac{\mu_0\Lambda}{4}\int j_s^2\,dA+\int j_sA_a\,dA,
\end{split}\label{eq:G}
\end{equation}
with area element $dA=2\pi r\,dr$. The three terms represent the magnetic energy of the self-field in all space, the kinetic energy of the supercurrent ($\int\tfrac12\mu_0\lambda^2j^2\,dV$ for a current uniform across $d$), and the coupling to the applied field. Setting the first variation of $\mathcal G$ with respect to $j_s(r)$ to zero yields Eq.~(\ref{eq:sheet}). Because $\mathcal G$ is a positive-definite quadratic functional plus a linear driving term, its stationary point is a unique global minimum. Note that Eq.~(\ref{eq:sheet}) is the purely inductive (Meissner--London) limit of Brandt's integral equation for a thin disc in an ac field~\cite{brandt1994a,brandt1994b,brandt1994c}.

The magnetic moment is $m=\pi a^3\int_0^1x^2j_s(x)\,dx$, giving susceptibility magnitude $|\chi|=|m|/(\pi a^2dH)$. Because all radial lengths scale with $a$, the normalized sheet susceptibility depends only on the single dimensionless parameter $\epsilon=\Lambda/a$:
\begin{equation*}
\frac{|\chi|}{\chiz}=F(\epsilon),\qquad \chiz=\frac{8a}{3\pi d}=\frac{4}{3\pi\delta},
\end{equation*}
where $\chiz$ follows from the ideal ($\epsilon=0$) sheet current, of magnitude $j_{s0}(x)=(4H/\pi)\,x/\sqrt{1-x^2}$~\cite{mikheenko1993,clem1994}.

We solve Eq.~(\ref{eq:sheet}) numerically by the Galerkin method~\cite{atkinson1997}: $j_s(r)$ is approximated as piecewise constant on concentric rings $R_{i-1}<r<R_i$ of width $w_i$ and mid-radius $r_i$, graded geometrically toward the rim $r=a$ (with the outermost ring width $10^{-7}a$--$10^{-9}a$ so that the edge boundary layer of width $\sim\Lambda$ is fully resolved), and the equation is imposed on the flux averaged over each ring. For the ring currents $I_i=j_s(r_i)w_i$ this gives
\begin{equation}
(\mathsf M+\mathsf L)\,I=-\Phi_a,\label{eq:discrete}
\end{equation}
where $\mathsf M_{ij}$ is the mutual inductance of rings $i$ and $j$ (the double average of $\mathcal M$ over their widths), $\mathsf L_{ii}=\mu_0\lambda^2\,2\pi r_i/(w_id)$ is the kinetic inductance of ring $i$, and $\Phi_{a,i}=\mu_0H\pi\langle r^2\rangle_i$ is the applied flux through it; each row states fluxoid conservation in one ring, and the solution minimizes the discretized energy $\mathcal G$, Eq.~(\ref{eq:G}). The ring--ring inductances are evaluated by Gauss--Legendre quadrature \cite{abramowitz1964}: three quadrature points per ring for well-separated rings, and six for adjacent rings or rings near the $r=0$ axis whose width is comparable to their radius. For self- and adjacent-ring integrals, the logarithmic singularity $\mathcal M(r,r')\simeq-\mu_0\sqrt{rr'}\ln|r-r'|$ is subtracted and integrated analytically over the ring widths \cite{davis1984}. Since $\mathsf L$ is proportional to $\Lambda$, $\mathsf L=\epsilon\mathsf D$ with $\mathsf D_{ii}=\mu_0\pi a\,r_i/w_i$, a single generalized eigenvalue problem $\mathsf Mv_k=\mu_k\mathsf Dv_k$ yields $F(\epsilon)$ for all $\epsilon$ simultaneously as a spectral sum of poles,
\begin{equation}
F(\epsilon)=\sum_k\frac{a_k}{1+\epsilon/\mu_k},\qquad\sum_ka_k=1,\label{eq:poles}
\end{equation}
with $a_k\propto(v_k^{\rm T}\Phi_a)^2/\mu_k>0$ for $v_k^{\rm T}\mathsf Dv_k=1$. This has the structure of Brandt's mode expansion~\cite{brandt1994c}, here evaluated with $\sim10^3$ eigenmodes rather than 20. The solver reproduces the ideal moment $|m(\lambda=0)|=\tfrac83a^3H$ to $10^{-8}$ and the weak-screening limit $F(\epsilon)\to3\pi/(32\epsilon)$ to $10^{-6}$, and agrees with Brandt's tabulated values~\cite{brandt1994c} to their stated precision of $2\times10^{-4}$; mesh refinement changes the screening deficit $1-F(\epsilon)$ by less than $0.1\%$ for all $\epsilon\ge5\times10^{-8}$.

In the pure 2D sheet model, the thickness $d$ enters only through Pearl's $\Lambda=2\lambda^2/d$ and the prefactor $\chiz\propto1/d$. As we show next, replacing $2\lambda^2/d$ by the effective length $\Leff$ of Eq.~(\ref{eq:Leff}) makes the sheet reproduce the exterior field of a film of finite thickness away from the rim. Rectangular films (Sec.~\ref{sec:rect}) are solved similarly by expressing the divergence-free sheet current via a stream function, $\mathbf j_s=\nabla\times(g\hat z)$, and minimizing the London energy on a 2D Cartesian mesh graded toward all four edges~\cite{brandt1995,brandt2005}.

\section{A semi-empirical closed-form susceptibility}\label{sec:model}

Our primary practical motivation is to construct a fast, easy to implement, yet accurate closed-form representation of $\chi(\lambda,d,a)$ that can be inverted directly by experimentalists to extract $\lambda(T)$ from whole-sample susceptibility measurements across all realistic film thicknesses and penetration depths.

Finite thickness $d$ modifies the 2D sheet response through three distinct physical mechanisms:

(a) \emph{Through-thickness variation of the current and vector potential not too close to the rim.} Locally, the film behaves as an infinite parallel slab, $|z|<d/2$, whose top and bottom surfaces experience tangential self-fields of equal magnitude and opposite sign, $\pm H_t = \pm j_s/2$, while the uniform perpendicular applied field contributes no tangential component. The tangential magnetic field inside the slab is therefore antisymmetric across the midplane, $H_r(z)=H_t\sinh(z/\lambda)/\sinh(d/2\lambda)$, vanishing at $z=0$, while the azimuthal current density $j_\phi(z)\propto\cosh(z/\lambda)$ flows in the same direction across the entire thickness. Integrating across $z\in[-d/2,d/2]$ gives $j_s=2H_t$ and a surface vector potential $A_\phi(\pm d/2)=-\mu_0\lambda H_t\coth(d/2\lambda)$. Outside the slab ($|z|>d/2$), the magnetic field is identical to that of a zero-thickness current sheet at the midplane $z=0$ carrying the same total sheet current $j_s$. Analytically continuing the exterior vector potential from the physical surfaces $|z|=d/2$ to the equivalent sheet plane $z=0$ via $\partial A_\phi/\partial|z|=-\mu_0H_t$ adds $+\mu_0H_td/2$, yielding Pearl's law $j_s=-(2/\mu_0\Leff)A_\phi(z=0)$ with $\Lambda$ replaced by the \emph{effective} screening length
\begin{equation}
\begin{split}
\Leff&=\lambda\left[\coth\frac{d}{2\lambda}-\frac{d}{2\lambda}\right]\\
&=\frac{2\lambda^2}{d}\left[1-\frac{d^2}{6\lambda^2}-\frac{d^4}{720\lambda^4}-\ldots\right]\quad(d<2\pi\lambda)\\
&=\lambda-\frac d2+O\big(e^{-d/\lambda}\big)\quad(d\gg\lambda).
\end{split}\label{eq:Leff}
\end{equation}
The expansion parameter for thin films is $(d/\lambda)^2$. Pearl's $\Lambda=2\lambda^2/d$ is the leading term; it corresponds to the thickness average of $\mathbf A$. The value of $\mathbf A$ at the surface exceeds this average because of the $\cosh(z/\lambda)$ profile, which adds $+d/6$, and continuing the exterior potential to the midplane subtracts $d/2$. The net leading finite-thickness shift is $\Leff-\Lambda \simeq -d/3$, which reduces $\Lambda$ by $1\%$ at $\lambda=4d$. Notably, $\Leff$ passes through zero at $d/\lambda=2.39936$ ($d\simeq2.40\lambda$) and becomes negative for thicker films ($d>2.40\lambda$), where supercurrents flow within skin layers of thickness $\sim\lambda$ near the top and bottom faces located at distance $d/2$ from the midplane. We use the exact closed form on the first line of Eq.~(\ref{eq:Leff}) for all calculations, never the truncated series.

(b) \emph{Finite-thickness correction to the ideal ($\lambda=0$) limit.} At $\lambda=0$, a disc of finite thickness has $|\chi|=(8a/3\pi d)\,E(\delta)$ with $E>1$. Its magnetic moment exceeds the thin-disc value $\tfrac83a^3H$ because the excluded flux must go around a rim of finite height $d$, and the edge singularity of the sheet current is cut off at a distance $\sim d$ from the rim. To leading order, $E\simeq1+(3/\pi)\,\delta\ln(3.55/\delta)$, which is equivalent to enlarging the disc radius by $(d/2\pi)\ln(3.55/\delta)$. The total $|\chi|\propto E/d$ still decreases with $d$; only its ratio to $8a/3\pi d$ grows.

(c) \emph{Three-dimensional current redistribution at the rim.} Within a radial distance $\sim d$ of the edge $r=a$, the current distribution is genuinely three-dimensional rather than a sheet current, producing a rim correction of relative order $\delta$.

Combining these three ingredients, we express the susceptibility magnitude $|\chi(\lambda,d,a)|$ in factored form,
\begin{equation}
\begin{split}
|\chi(\lambda,d,a)|&=\frac{8a}{3\pi d}\;\mathcal F\;\big[1-2\delta\,h(\lambda/d)\big],\\
\mathcal F&=\begin{cases}F(\Leff/a),&\Leff>0,\\ E(|\Leff|/a),&\Leff\le0,\end{cases}
\end{split}\label{eq:model}
\end{equation}
and determine its empirical parameters by fitting to the full 3D numerical solutions of Sec.~\ref{sec:methods}. Exact analytical asymptotes fix the functional structure at both ends, while the intermediate crossover constants are fitted.

\textit{Sheet function} $F(\epsilon)$: the normalized susceptibility of a zero-thickness disc as a function of $\epsilon=\Lambda/a$, evaluated in Eq.~(\ref{eq:model}) at $\epsilon=\Leff/a$, is represented across all $\epsilon>0$ by
\begin{equation}
\begin{split}
\frac1{F(\epsilon)}-1=\epsilon\Big[&\frac3\pi\ln\Big(1+\frac{c_2}{\epsilon}\Big)+\frac{32}{3\pi}\\
&+\frac{c_3}{1+c_4\epsilon}+\frac{c_5\,\epsilon}{(1+c_6\epsilon)^2}\Big],
\end{split}\label{eq:F}
\end{equation}
with $c_2=0.0401744$, $c_3=0.8898535$, $c_4=8.093975$, $c_5=20.06212$, and $c_6=94.44745$. Equation~(\ref{eq:F}) reproduces the exact weak-screening asymptote $F(\epsilon)\to 3\pi/(32\epsilon)$ as $\epsilon\to\infty$ and the strong-screening logarithmic law $1-F(\epsilon)\simeq (3/\pi)\epsilon\ln(3.571/\epsilon)$ as $\epsilon\to0$, whose coefficient $3/\pi$ follows from first-order perturbation in the kinetic energy and agrees with Brandt's numerical asymptote~\cite{brandt1994b}; the matched asymptotics of Ref.~\cite{dorsey1995} give the same form with the coefficient $2\sqrt2/\pi$, $6\%$ lower. Equation~(\ref{eq:F}) is accurate to $0.02\%$ in $F$ and $0.15\%$ in the deficit $1-F$ over $2\times10^{-7}\le\epsilon\le200$, and remains asymptotically exact beyond both ends.

\textit{Effective length} $\Leff$, Eq.~(\ref{eq:Leff}): for $\lambda<d/2.3994$, $\Leff\le0$; in this thick-film regime, the top and bottom skin layers screen like an ideal disc of effective thickness $d_{\rm eff}=2|\Leff|\simeq d-2\lambda$, which enters the second branch of $\mathcal F$ through the effective aspect ratio $\delta_{\rm eff}=d_{\rm eff}/(2a)=|\Leff|/a$.

\textit{Finite-thickness ideal limit} $E(\delta)=|\chiz(d)|/(8a/3\pi d)$:
\begin{equation}
\begin{split}
E(\delta)&=\frac{3\pi}{4}(\delta+q),\\
q&=\frac{4}{3\pi}+\frac{2}{3\pi}\\
&\quad\times\tanh\!\Big[\frac6\pi\,\delta\ln\Big(1+\frac{0.17760}{\delta}\Big)+1.00822\,\delta\Big],
\end{split}\label{eq:E}
\end{equation}
which follows the functional form proposed by Brandt~\cite{brandt2001} and is accurate to $0.05\%$ for $\delta\le0.017$ and $0.3\%$ up to $\delta=0.25$. For thin discs ($\delta\ll1$), Eq.~(\ref{eq:E}) expands as $E(\delta)\simeq1+(3/\pi)\,\delta\ln(3.55/\delta)$; the finite-thickness enhancement $E-1$ reaches $1.1\%$ at $\delta=1.5\times10^{-3}$ and $8.8\%$ at $\delta=0.017$. In Eq.~(\ref{eq:model}), $E$ is evaluated at $\delta_{\rm eff}=|\Leff|/a$, which reduces to the geometric aspect ratio $\delta=d/(2a)$ at $\lambda=0$ ($\Leff=-d/2$) and smoothly approaches $E(0)=1=F(0)$ as $\Leff\to0$.

\textit{Rim function} $h(\lambda/d)$, universal for $\delta\le0.05$: defining the reduced variable $w=2.3994\,\lambda/d$ (so that $w=1$ at $\Leff=0$) and $v=w-1$,
\begin{equation}
h=\begin{cases}\dfrac{w(1.15083-0.26951\,w)}{1+0.50433\,w^{7.3113}},&w<1,\\[8pt]
\dfrac{0.55798}{1+2.11811\,v^{0.84584}+0.59756\,v^{1.75785}},&w\ge1,\end{cases}\label{eq:h}
\end{equation}
where the two branches are bridged by linear interpolation over $|v|<0.04$ ($0.96<w<1.04$). At small $\lambda/d$, $h$ grows linearly as $h\simeq2.761\,\lambda/d$ (corresponding to radial flux penetration shrinking the effective rim radius by $\propto\lambda$), reaches a maximum of $0.682$ near $\lambda/d=0.34$, and falls off as $v^{-1.76}$ (approximating the $\sim(d/\lambda)^2$ decay over the fitted range) for $\lambda\gg d$.

\textit{Overall accuracy.} Benchmarked against the full 3D numerical solutions across the entire range of $\lambda$, Eq.~(\ref{eq:model}) is accurate to $0.07\%$ for $\delta\le1.7\times10^{-3}$, $0.08\%$ at $\delta=0.017$, and $0.11\%$ at $\delta=0.05$. Where higher precision is required, the sheet integral equation or 3D problem can be solved directly.

\section{When is a film two-dimensional?}\label{sec:2d3d}

\begin{figure}[tbh]
\includegraphics[width=8.4cm]{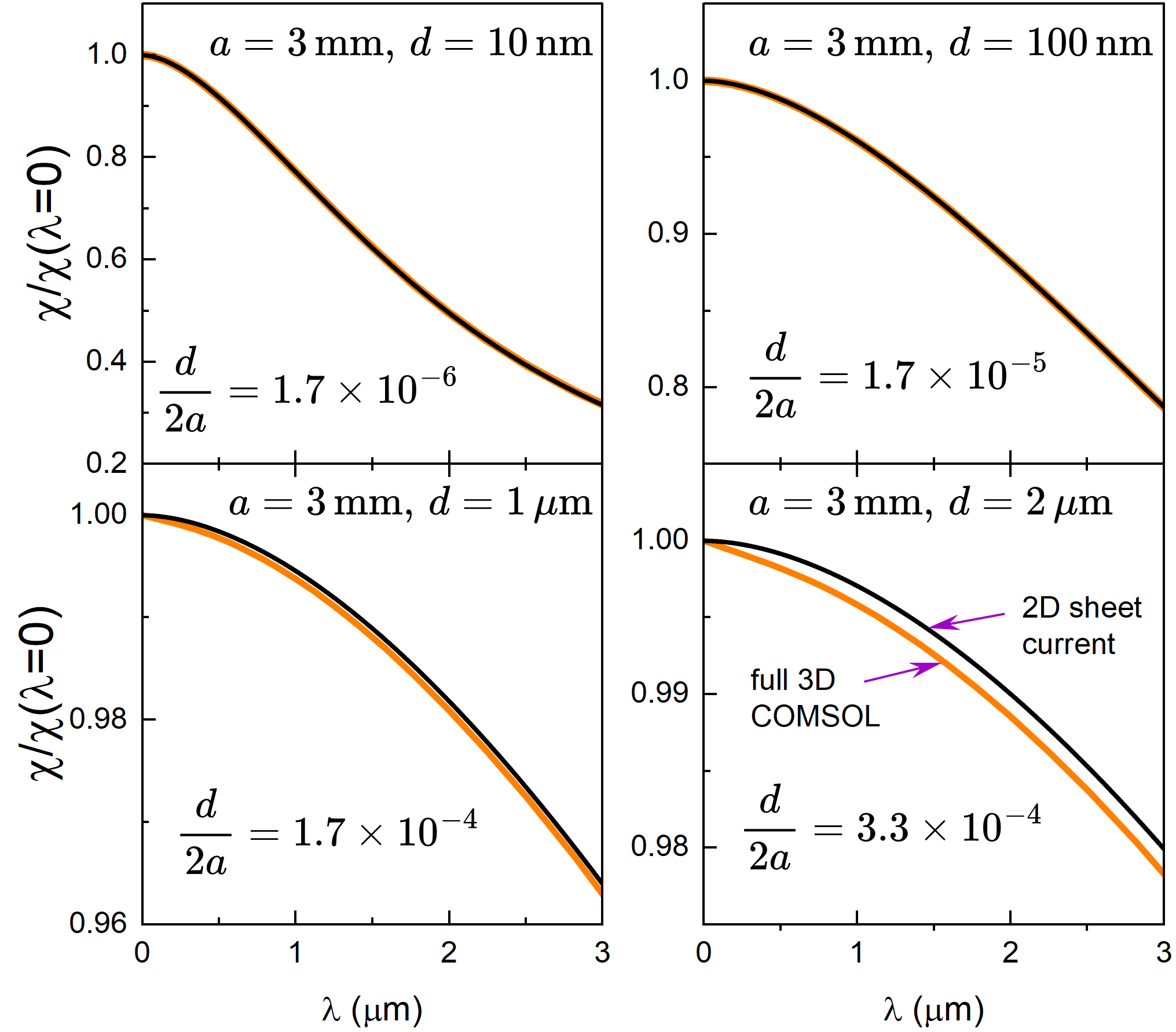}
\caption{Normalized susceptibility $\chi/\chi(\lambda=0)$ of a disc with $a=3$~mm as a function of $\lambda$: pure 2D sheet-current model with $\Lambda=2\lambda^2/d$ (black curves) vs. full 3D finite-element solution (COMSOL, orange curves) for $d=10$~nm, $100$~nm, $1$\um, and $2$\um\ ($\delta=d/(2a)=1.7\times10^{-6}$ to $3.3\times10^{-4}$).}
\label{fig:2d3d}
\end{figure}

For a superconducting film in a perpendicular magnetic field, ``two-dimensional'' has (at least) three distinct physical meanings, governed by three different length-scale ratios:

\textit{1. Electrodynamic 2D regime (sheet vs. volume current).}
The 2D sheet approximation requires the current distribution to vary laterally on scales much larger than $d$, so that locally the film acts as a planar slab with tangential fields $\pm H_t$ on its faces and the exterior field senses only the integrated sheet current $j_s$. Figure~\ref{fig:2d3d} compares the pure 2D sheet model (using the unshifted Pearl length $\Lambda=2\lambda^2/d$, normalized to its $\lambda=0$ limit) against the full 3D COMSOL finite-element solutions for a disc of radius $a=3$~mm and four thicknesses. For $\delta=d/(2a)\le1.7\times10^{-5}$ ($d\le100$~nm), the 2D sheet and 3D curves are virtually indistinguishable over $\lambda\le3$\um. At $\delta=1.7\times10^{-4}$ ($d=1$\um), they separate visibly, and at $\delta=3.3\times 10^{-4}$ ($d=2$\um) the pure sheet model misses $\sim8\%$ of the susceptibility deficit $1-|\chi|/|\chi(\lambda=0)|$ at $\lambda=3$\um\ and $\sim30\%$ of it at $\lambda=1$\um. Because $\lambda(T)$ in experiments is extracted directly from this deficit, the error in $\lambda$ is substantial when the pure 2D sheet formula is applied to micron-scale films. We conclude that the bare 2D Pearl sheet model is quantitatively reliable for $\delta \lesssim 10^{-4}$ and $\lambda \gtrsim 4d$.

\textit{2. Screening crossover (Pearl length vs. lateral sample size).}
The ratio $\epsilon=\Lambda/a$ separates strong diamagnetic screening ($\Lambda\ll a$, where $1-F(\epsilon)\propto\epsilon\ln(C/\epsilon)$ is controlled by the edge peak) from weak screening ($\Lambda\gg a$). In the weak-screening limit, $F(\epsilon)\to 3\pi/(32\epsilon)$ so that $|\chi|\to a^2/(8\lambda^2)$, which measures $\lambda^{-2}$ directly and is independent of both the film thickness $d$ and the ideal-screening normalization $\chiz$.

\textit{3. Thermodynamic 2D regime (thickness vs. coherence length).}
When $d\lesssim\xi(T)$, the superconducting order parameter is uniform across the thickness and thermal phase fluctuations become two-dimensional. This modifies the physical penetration depth $\lambda(T)$ itself---through surface scattering, thickness-induced $T_c$ suppression, and the Berezinskii--Kosterlitz--Thouless (BKT) renormalization of the 2D superfluid stiffness $\Lambda^{-1}(T)$ by bound vortex--antivortex pairs~\cite{berezinskii1972,kosterlitz1973,nelson1977,beasley1979,halperin1979,hebard1980,ambegaokar1980}---but it does not alter the macroscopic Maxwell--London electrodynamics: Eq.~(\ref{eq:model}) remains valid with whatever renormalized $\lambda(T)$ the film possesses, provided the current--field relation remains local (i.e., Pippard coherence length $\xi_0\ll\lambda$ or mean free path $\ell\ll\lambda$).

\section{Rectangular samples}\label{sec:rect}

Most experimental thin-film samples are rectangular platelets rather than circular discs. To make our closed-form disc model directly applicable to realistic sample shapes, we solved the 2D sheet-current integral equation numerically for rectangular films of lateral dimensions $w\times l$ ($w\le l$) and thickness $d$ on a 2D mesh graded toward the edges~\cite{brandt1995,brandt2005}, covering aspect ratios $1\le l/w\le16$ and the full range of $\Lambda$. We then fitted the normalized rectangular response $|\chi(\Lambda)|/|\chi(0)|$ to the normalized disc sheet function $F(\Lambda/a_r)$ of Eq.~(\ref{eq:F}) using a single effective disc radius $a_r(w,l)$ as the mapping parameter, obtaining
\begin{equation}
a_r=w\left[0.735-0.222\left(\frac wl\right)^{1.32}\right],\qquad w\le l.\label{eq:ar}
\end{equation}
At each aspect ratio $l/w$, $a_r$ was chosen to minimize the maximum relative error in the inverted $\lambda$ across the experimentally relevant window $10^{-4}\le\Lambda/w\le3$, and Eq.~(\ref{eq:ar}) reproduces these optimal radii to within $0.5\%$. Using $a_r$ in $F(\Lambda/a_r)$, the error in the extracted $\lambda$ remains within $\pm1.1\%$ over six decades of screening, $10^{-5}\le\Lambda/w\le30$. For a square film ($w=l$), Eq.~(\ref{eq:ar}) gives $a_r=0.513\,w$; for a long strip ($w/l\to0$), $a_r\to0.735\,w$; for a typical $3\times4$~mm$^2$ rectangular substrate ($w=3$~mm, $l=4$~mm)~\cite{moirangthem2026b}, $a_r=1.75$~mm.

Other common choices for mapping a rectangle onto a disc perform significantly worse: using the inscribed radius $w/2$, the equal-area radius $(wl/\pi)^{1/2}$, or the radius of equal ideal susceptibility produces errors in the inverted $\lambda$ of $2\%$, $6\%$, and $3\%$ for a square ($l/w=1$), and $16\%$, $27\%$, and $8\%$ for an elongated rectangle with $l/w=4$. Naturally, no single effective radius can be exact for all $\Lambda/w$, because the weak-screening limit ($\Lambda\gg w$) is governed by the global bulk current circulation whereas the strong-screening limit ($\Lambda\ll w$) is governed by the perimeter edges and corners.

Importantly, the mapping $a\to a_r$ transfers the \emph{normalized shape} $|\chi(\lambda)|/|\chi(0)|$ of the response, not its absolute amplitude. The ideal ($\lambda=0$, $\delta\to0$) susceptibility of a rectangular film is $|\chiz^{\rm rect}|=|m_0|/(wldH)$ with $|m_0|=(\pi/4)\,w^2lH\,[1-0.548\,w/l+0.126\,(w/l)^2]$ (accurate to $0.15\%$), or as given by the empirical fit of Chen \emph{et al.}~\cite{chen2010} (accurate to $0.1\%$), rather than $8a_r/(3\pi d)$. One should also bear in mind that Eq.~(\ref{eq:ar}) was derived within the 2D sheet model and has been established only for $d/(2a_r)\lesssim10^{-4}$. Within that range, the largest error in $\lambda$, $1.0\%$, occurs for the square, and it is at most $0.5\%$ for $2\le l/w\le4$.

\section{Calibration of whole-sample measurements}\label{sec:calib}

Finally, we formulate the practical calibration procedure for whole-sample susceptibility measurements, such as frequency-domain tunnel-diode resonator (fTDR), microwave cavity perturbation, or global ac susceptometry. If the raw measured signal is $S(T)$ (for example, the resonator frequency $f(T)$), the shift measured from the transition temperature $T_c$, $\Delta S(T) = S(T_c)-S(T)$, is directly proportional to the susceptibility magnitude (since $\chi(T_c)=0$),
\begin{equation}
\Delta S(T)=G\,|\chi(T)|,\label{eq:dS}
\end{equation}
with an unknown apparatus calibration constant $G$.

In bulk crystals where $\lambda(0)$ is much smaller than the effective sample dimension $\Reff$~\cite{prozorov2000,prozorov2021}, $(1-N)|\chi(T\to0)|\simeq 1-\lambda(0)/\Reff\approx 1$, so normalizing $\Delta S(T)$ by its base-temperature value $\Delta S(0)$ directly yields $|\chi(T)|/\chiz$, and low-temperature changes $\Delta\lambda(T)=\lambda(T)-\lambda(0)$ are proportional to $S(T)-S(0)$ without requiring precise knowledge of $\lambda(0)$.

In thin and ultrathin films in a perpendicular field, by contrast, the Pearl length $\Lambda(0)=2\lambda^2(0)/d$ can be a substantial fraction of the lateral size $a$ even at $T\to0$. Consequently, the base-temperature susceptibility $|\chi(\lambda(0))|$ is significantly smaller than the ideal value $|\chi(\lambda=0)|$ (for example, $|\chi(\lambda(0))|/|\chi(0)|\approx0.62$ for a $d=10$~nm film with $\lambda(0)\approx1.14$\um\ and $a_r=1.75$~mm~\cite{moirangthem2026b}), and the curve $\chi(\lambda)$ is strongly nonlinear. Dividing $\Delta S(T)$ by its extrapolated zero-temperature value $\Delta S(T\to0)=S(T_c)-S(T\to0)$ eliminates the unknown instrumental constant $G$:
\begin{equation}
\frac{\Delta S(T)}{\Delta S(T \to 0)}=\frac{\chi(T)}{\chi(T \to 0)} = \frac{\chi(\lambda(T))}{\chi(\lambda(0))}.\label{eq:calib_ratio}
\end{equation}
To invert the nonlinear function $\widetilde\chi(\lambda)\equiv|\chi(\lambda,d,a)|/|\chi(0;d,a)|$ calculated from Eq.~(\ref{eq:model}) (using $a=a_r$ from Eq.~(\ref{eq:ar}) for a rectangular film), the experimental ratio in Eq.~(\ref{eq:calib_ratio}) is multiplied by the calculated base-temperature deficit factor $\widetilde\chi(\lambda(0))$, where $\lambda(0)$ is taken from an independent measurement or estimate (such as aluminium coating~\cite{prozorov2000b}, optical/transport Drude weight, or $\mu$SR):
\begin{equation}
\widetilde\chi_{\rm exp}(T) \equiv \frac{\Delta S(T)}{\Delta S(T \to 0)}\,\widetilde\chi(\lambda(0)) = \widetilde\chi(\lambda(T)).\label{eq:calib_invert}
\end{equation}
Numerically inverting $\widetilde\chi(\lambda)=\widetilde\chi_{\rm exp}(T)$ at each temperature $T$ then directly yields the intrinsic $\lambda(T)$ and the normalized superfluid density $\rho_s(T)=[\lambda(0)/\lambda(T)]^2$~\cite{moirangthem2026,moirangthem2026b}. Notice that in the 2D sheet regime ($\delta\lesssim10^{-4}$, $\lambda\gtrsim4d$), where $\widetilde\chi=F(\Lambda/a)$ depends on $\lambda$ and $d$ solely through $\Lambda=2\lambda^2/d$, the measurement determines $\Lambda(T)$ once $\Lambda(0)$ is fixed. If the reference is $\Lambda(0)$ itself (for example, from a two-coil measurement), the film thickness drops out of $\rho_s(T)=\Lambda(0)/\Lambda(T)$ entirely and is needed only to convert $\Lambda$ into $\lambda$. If instead $\lambda(0)$ is the reference, an error in $d$ enters through $\Lambda(0)=2\lambda^2(0)/d$ and propagates into $\rho_s(T)$.

\section{Conclusions}

In conclusion, two independent full 3D Meissner--London numerical solutions for a superconducting disc in a perpendicular magnetic field were used to establish the quantitative range of validity of the 2D Pearl sheet-current model: $\delta=d/(2a)\lesssim10^{-4}$ and $\lambda\gtrsim4d$. To eliminate the need for tedious 3D numerical calculations in data analysis, we constructed a semi-empirical closed-form expression for $|\chi(\lambda,d,a)|$, Eq.~(\ref{eq:model}), constrained by exact analytical limits and accurate to $0.1\%$ for all $\lambda$ and aspect ratios $d/(2a)\le0.05$. For rectangular films, we determined a phenomenological effective disc radius $a_r(w,l)$, Eq.~(\ref{eq:ar}), that maps the normalized rectangular sheet response onto the disc curve to within $1.1\%$ in the extracted $\lambda$ across six decades of $\Lambda/w$. Finally, we formulated the complete whole-sample calibration protocol, Eqs.~(\ref{eq:calib_ratio})--(\ref{eq:calib_invert}), enabling reliable extraction of the intrinsic London penetration depth $\lambda(T)$ and superfluid density $\rho_s(T)$ from global susceptibility measurements such as conventional ac susceptometry, microwave cavity perturbation, and planar frequency-domain tunnel-diode resonator (fTDR) techniques.

\begin{acknowledgments}
This work was supported by the U.S. Department of Energy, Office of Science, National Quantum Information Science Research Centers, Superconducting Quantum Materials and Systems Center (SQMS), under Contract No. 89243024CSC000002, and by DOE BES, Materials Science and Engineering Division. Ames Laboratory is operated for the U.S. DOE by Iowa State University under contract DE-AC02-07CH11358. 
\end{acknowledgments}

%

\end{document}